\documentclass[prd,nofootinbib,preprint,superscriptaddress]{revtex4}

\usepackage{amsmath, amssymb, amsthm, graphicx, epsfig, fancyhdr,epsfig, slashed, mathrsfs}

\usepackage{tikzsymbols}
\usepackage{natbib}
\usepackage{float}

\usepackage{mathtools} 

\usepackage{tikz,xcolor,hyperref}

\definecolor{lime}{HTML}{A6CE39}
\DeclareRobustCommand{\orcidicon}{
	\begin{tikzpicture}
	\draw[lime, fill=lime] (0,0) 
	circle [radius=0.2] 
	node[white] {{\fontfamily{qag}\selectfont \tiny ID}};
	\draw[white, fill=white] (-0.0625,0.095) 
	circle [radius=0.007];
	\end{tikzpicture}
	\hspace{-2mm}
}

\foreach \x in {A, ..., Z}{\expandafter\xdef\csname orcid\x\endcsname{\noexpand\href{https://orcid.org/\csname orcidauthor\x\endcsname}
			{\noexpand\orcidicon}}
}

\newcommand{\be}{\begin{equation}}
\newcommand{\ee}{\end{equation}}
\newcommand{\bea}{\begin{eqnarray}}
\newcommand{\eea}{\end{eqnarray}}

\def\({\left(}
\def\){\right)}
\def\[{\left[}
\def\]{\right]}

\newcommand{\ag}[1]{\textcolor{blue}{[Anish: #1]}}

\begin{document}

%\title{Confinement Conditions and $\beta$-functions in String-inspired \\ Infinite Derivative Non-local  Gauge Theories}

\title{Generalising Axion-like particle as the curvaton: \\ \it{sourcing primordial density perturbation and non-Gaussianities}}

\author{Anish Ghoshal\orcidA{}}
\email{anish.ghoshal@fuw.edu.pl}
%\affiliation{INFN Rome, Italy}
\affiliation{Institute of Theoretical Physics, Faculty of Physics, University of Warsaw, ul. Pasteura 5, 02-093 Warsaw, Poland}

\author{Abhishek Naskar\orcidB{}}
\email{abhiatrkmc@gmail.com}
\affiliation{Department of Physics, Indian Institute of Technology Bombay, \\ Mumbai 400076, India}

%\author{Alessio Notari\orcidC{}}
%\email{notari@fqa.ub.edu}
%\affiliation{Departament de F\'isica Fondamental i Institut de Ci\`encies del Cosmos, Universitat de Barcelona, Mart\'i i Franqu\`es 1, 08028 Barcelona, Spain.}

\begin{abstract}

We investigate the non-perturbatively generated axion-like particle (ALP) potential, involving fermions in the dark sector that couple to the ALP, in an early cosmological inflationary  stage with the ALP being a spectator field. The potential here deviates from the standard cosine nature due to the presence of the two fermion masses $m_u$ and $m_d$ which couple to the ALP. The ALP is a spectator field during inflation but it starts to oscillate and dominates the energy density of the universe after inflation ends, thereby sourcing isocurvature perturbations, while standard curvature fluctuations form the inflaton are assumed to be sub-dominant. Subsequently the ALP decays converting the isocurvature perturbations to adiabatic perturbations thereby acting as the origin of the primordial density perturbations. We identify the parameter space involving the axion decay constant $f_a$, scale of confinement $\Lambda$, ALP mass $m$ and the masses of the fermions, $m_u$ and $m_d$ where it can satisfactorily behave as the curvaton and source the observed primordial density perturbation. We also predict local non-Gaussianity signals for bi-spectrum and tri-spectrum $f_{NL}$ and $g_{NL}$, as  a function of the ratio $m_u/m_d$, which are within the allowed range in the latest Planck observations and are detectable with future observations. Particularly we observed that the value of $f_{NL}$ and $g_{NL}$ are dependent on the ratio of $m_u$ and 
$m_d$: $f_{NL}$ is more or less positive for all scenarios except $m_u = m_d$ and $g_{NL}$ is always positive irrespective of the ratio between $m_u$ and 
$m_d$. The results of our analysis in the limit $m_u = m_d$ resembles vanilla curvaton scenario while in the limit $m_u \gg m_d$ resembles pure axion cosine potential.

\end{abstract}

\maketitle

\section{Introduction}

Even before the cosmological scales starts entering the Hubble horizon of the universe, the primordial curvature perturbation $\zeta$ is already present, at least few Hubble times before this. It must be during this era, and particularly on those scales, the Fourier components $\zeta (k)$ (k being momentum) are  time-independent. They set the  initial condition for the large-scale structure (LSS) formation in the Universe \cite{book}. 
Therefore, $\zeta(k)$ is determined, leading to a direction of model-buildings in theoretical cosmology that lead to the origin of $\zeta$.

Usually it is believed that the generation of $\zeta(k)$  presumably starts at the time of horizon exit during inflation (that is, when $ k=aH\equiv \dot a$ where $a(t)$ is the FRW metric scale factor of the universe) when the vacuum fluctuation of one or more 
scalar (or vector  \cite{dklr}) fields takes the form of
a classical  perturbation. In a 
slow-roll inflationary model of early universe with a single-field $\phi_I$,
$\zeta$ is generated by the perturbation $\delta\phi_{I}$, which means that $\zeta$ is generated only at the time of horizon exit, and remains constant afterwards. On the other hand, another alternative and well-motivated scenario is the curvaton scenario \cite{ourcurv}, in which case $\zeta$ is now instead generated by the perturbation $\delta \sigma$ of 
 a `curvaton' field $\sigma_c$, that has practically
no effect, or almost remains frozen during inflation. This means that $\sigma_c$ generates $\zeta$ only and only when its energy density becomes significant, mostly when inflation has ended. One may also have mixed inflaton-curvaton
scenario where both $\delta\phi_I$ and $\delta\sigma_c$ contributes significantly to $\zeta$ \cite{othercurv}, $\zeta$ be generated during  multi-field inflation, or via  a `modulating' field that induces an effective mass or coupling to be inhomogeneous \cite{book}.

In this paper, we will stick to purely the curvaton scenario however assuming single-field inflationary model to hold true. In our case we will look at make axion or axion-like particle as the curvaton. In context to the Standard Model (SM) of particle physics, something well-known as the The Strong CP problem is one of the unsolved problems; and it strongly recommends to go beyond the SM (BSM). The so-called Peccei-Quinn (PQ) mechanism~\cite{Peccei:1977hh,Peccei:1977ur}, with
predictions of a light pseudo-Nambu-Goldstone boson (pNGB), famously called the QCD axion, dynamically addresses the strong CP phase $\bar{\theta}_{\rm QCD}$ to be zero~\cite{Weinberg:1977ma,Wilczek:1977pj}. 
Non-perturbative effects of QCD, under strong confining dynamics generates QCD axion mass which must be lighter than ${\cal O}(10)$ meV to satisfy the current astrophysical observational 
bounds~\cite{Hamaguchi:2018oqw,Beznogov:2018fda,Leinson:2019cqv}. (see Refs.\,\cite{Jaeckel:2010ni,Ringwald:2012hr,Arias:2012az,Graham:2015ouw,Marsh:2015xka,Irastorza:2018dyq, DiLuzio:2020wdo} for reviews).

On the other hand, going beyond the standard QCD axion paradigm, we also envisage several Axion-like
particles (ALPs) that are also light gauge-singlet pseudoscalar bosons that couple weakly to the
Standard Model (SM) and generically appear as the pseudo-Nambu-Goldstone boson (pNGB), particularly
in theories with a spontaneously broken global $U(1)$ symmetry, and are motivated from various string
theory constructions \cite{Arvanitaki:2009fg}. ALPs could solve some other open questions in the
SM, such as the hierarchy problem of the Standard Model via the relaxion
mechanism~\cite{Graham:2015cka} responsible for inflationary cosmology~\cite{Freese:1990rb,
Adams:1992bn, Daido:2017wwb}, be non-thermal dark matter (DM) candidates~\cite{Preskill:1982cy,
Abbott:1982af, Dine:1982ah}, account for the dark energy in the universe~\cite{Jain:2004gi,
Kim:2009cp, Kim:2013jka, Lloyd-Stubbs:2018ouj}, and baryogenesis and
leptogenesis~\cite{Daido:2015gqa, DeSimone:2016bok} and recently considered in context to
non-standard cosmology with axion-diven kination
\cite{Co:2019jts,Co:2019wyp,Co:2020jtv,Harigaya:2021txz,Co:2021qgl}. \footnote{Recently, a minimal
extension involving SM supplemented with heavy neutrinos and a complex scale whose phase is the axion and the real part generates heavy neutrino masses have been proposed ~\cite{Salvio:2015cja,
Ballesteros:2016euj, Ballesteros:2016xej, Ema:2016ops, Salvio:2018rv, Gupta:2019ueh}.}.
%\ag{Check if all citations are done.}

In order to describe the scalar field as ALPs their coupling to the SM particles is suppressed by inverse powers of the $U(1)$ symmetry breaking energy scale $f_{\phi}$ (also known as the ALP decay constant). This energy scale is usually identified as the vacuum expectation value (VEV) of the SM-singlet complex scalar field $\Phi$, i.e. $\langle \Phi\rangle=f_{\phi}/\sqrt{2}$, quite larger the electroweak scale $v_{\rm ew}\simeq 246.2$ GeV in order to evade current experimental and observational limits~\cite{Jaeckel:2010ni, Irastorza:2018dyq}. The $\Phi$-field can be expressed as:
\begin{align}
\Phi(x) \ = \ \frac{1}{\sqrt 2}\left[f_{\phi}+\sigma(x)\right]e^{i\phi(x)/f_{\phi}} \, .
\label{eq:phix}
\end{align}
%\textcolor{red}{The notation has to be fixed, $\phi$ was already used for the inflaton.} \ag{Done.} 
Modulus $\phi$ of the $\Phi$-field receives a large mass term $m_\phi\sim f_{\phi}$, and the angular part $a$ becomes the pNGB that acquires a much smaller mass $m_a$ from an explicit low energy $U(1)$-breaking non-perturbative effects. Therefore, as usually done for any effective-field-theory (EFT) purposes for the low-energy phenomenology of ALPs, the heavier modulus part $\phi$ can be safely integrated out, and we are left with only independent parameters as the mass of the ALP $m$ and the decay constant $f_{\phi}$.
In presence of a dark sector with matter content with at least two fermions, let us call them up-type
fermion and down-type fermions\footnote{In pure QCD scenario, this corresponds to up and down type
quarks in QCD which couples to the axion, however in our case we will consider generic dark sector and consider an ALP
scenario with two dark fermions that couple to it with masses $m_u$ and $m_d$ respectively.}, the axion
potential does not show the pure cosine behavior but instead slightly modified as we will show in
Eqn. \eqref{main-pot} \cite{DiLuzio:2020wdo,GrillidiCortona:2015jxo}. Most of the early works
considering axion or ALP as the curvaton considered the radial part of the ALP field or small field
regimes of the phase part as the curvaton and studied the generation of primordial density
fluctuations as well as generation of Primordial Blackholes and secondary Gravitational Waves
\cite{Kawasaki:2011pd,Kawasaki:2012wr,Kawasaki:2012gg,
Ando:2017veq,Ando:2018nge,Kawasaki:2021ycf,Sasaki:1995aw}. Very recently in Ref.
\cite{Kobayashi:2020xhm} the non-perturbatively generated cosine part of axion was considered to be
the curvaton and investigated the parameter range where the axion may source the correct density
perturbations as well as predictions in non-Gaussianity signals. In this paper we will study such an
ALP coupled to two dark fermions with masses $m_u$ and $m_d$ in early universe. Particularly, we will
consider the ALP as a spectator field during inflation (with sub-dominant energy density compared  to
the inflaton) and then starts to oscillate and sources density perturbation acting as the curvaton.
What we will see is that various combinations of $m_u$ and $m_d$ masses that the ALP couples to we
actually we are able to achieve limits behave like quadratic potentials (like the vanilla curvaton)
on one hand and cosine potential on the other hand. The results of the respective curvaton scenarios
are thus obtained there-in in the same ALP model but due to various choices of the matter content and
confining dynamics that couples to the ALP. This will result in detectable predictions in the power
spectrum, bi-spectrum and tri-spectrum in CMB.

\textit{The paper is organized as follows:} in section II, we give a brief overview of the ALP model, in sections III and IV we study ALP as spectator field generating primordial curvature perturbations later on after inflation ends, and also discuss non-Gaussinaities. In section V, we show the results of our investigation and we end in section VI with a generalized discussion and the salient features of the ALP as the curvaton that have been manifest.

\medskip

\section{The axion setup}

%\textcolor{red}{The notation here is incoherent, in the previous section $a$ was the axion field, $\phi$ was the inflaton...}
%\ag{Fixed.}

%{\it The axion setup.}---
We consider a pseudo-Nambu goldstone boson (PNGB) of some global U(1) symmetry that is
spontaneously broken at an energy scale~$f_{\phi}$; we refer to this field as
the axion and $f_{\phi}$ the axion decay constant. 
We also assume the U(1) to be explicitly broken due to the axion
being coupled to a new gauge force (not QCD) that becomes strong at low
energies, yielding a periodic axion potential \cite{DiLuzio:2020wdo,GrillidiCortona:2015jxo}.
Considering the periodicity to be governed by $f_{\phi}$, 
we write the axion Lagrangian as 
\begin{equation}
\mathcal{L} = -\frac{1}{2} \partial_\mu \phi \partial^\mu \phi
- V(\phi)
%+ \frac{ \kappa }{8 \pi } \frac{\phi}{f_{\phi}} F_{\mu \nu} \tilde{F}^{\mu \nu }.
\label{eq:L}
\end{equation}

%\ag{write the total potential}

where the behavior of axion potential V($\phi$) can be written as,

\begin{equation}\label{main-pot}
    V(\phi) = m_{\pi}^2 f_{\pi}^2 \left[1-\sqrt{1-\frac{4 m_u m_d}{(m_u+m_d)^2}\sin^2\left(\frac{\phi}{2 f_{\phi}}\right)}\right] 
\end{equation}

where, $m_u$ and $m_d$ are masses of two light quarks, $m_{\pi}$ and $f_{\pi}$ are the mass and the decay constant of a "dark" pion, 
and $f_{\phi}$ is a constant. 

\begin{figure}[H]
    \centering
      \includegraphics[width=0.6\textwidth]{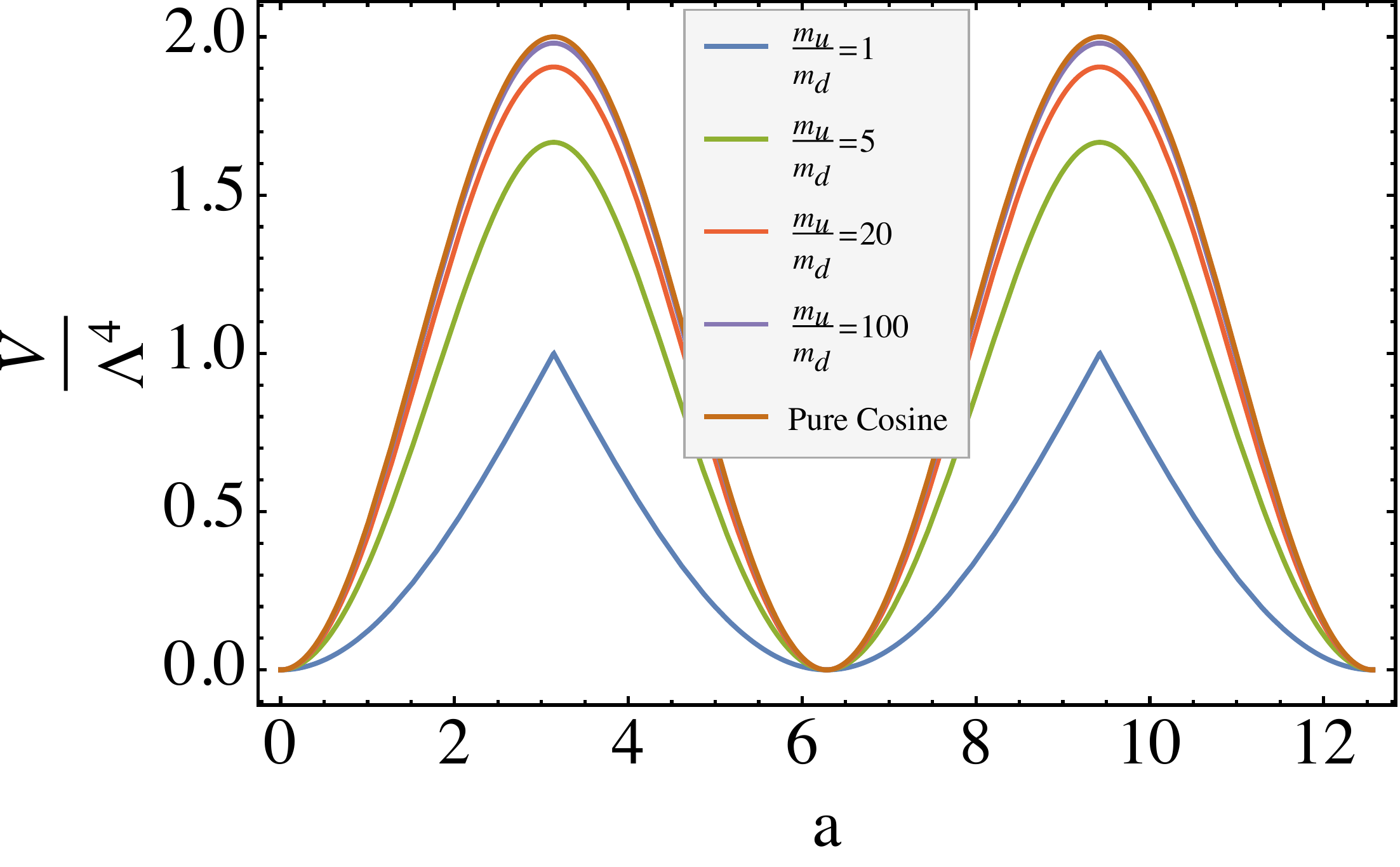}    
    \caption{\it  Plot of V(a) versus a (in arbitrary units) for various limits of $m_u$ and $m_d$.}
    \label{fig:potential}
\end{figure}

The potential \eqref{main-pot} depends on the ratio $\frac{m_u}{m_d}$
and in two extreme limits behaves as,

\begin{equation} 
V(\phi) = \Lambda^4 \frac{m_d}{m_u}\left[1-\cos\left(\frac{\phi}{f_{\phi}}\right)\right] 
             \mathrm{for}\, \, \, m_u \gg m_d
\end{equation}            

\begin{equation}\label{pot:mu=md}
V(\phi) =  \Lambda^4 \left[1-\cos\left(\frac{\phi}{2 f_{\phi}}\right)\right] 
             \mathrm{for}\, \, \, m_u = m_d
\end{equation}

From Fig \ref{fig:potential} we can see that $m_u=m_d$ limit behaves very similar to as vanilla
curvaton scenario (since below the cusp $V''(\phi)$ is always positive )and when $m_u\gg m_d$ it behaves as cosine like scenario. Here we have introduced the parameter and set it to
$\Lambda^4 = m_{\pi}^2 f_{\pi}^2 = m^2 f_{\phi}^2$. %Also, in \eqref{eq:L}we have added a pseudoscalar coupling to photons (or hidden photons). 
%the axion is written as a dimensionless angle~$a$
%with periodicity $a \cong a + 2 \pi$,
%and 

The zero-temperature axion mass is given by \footnote{Usually, below some strong coupling scale~$\Lambda$
($ <  f_{\phi}$), we consider the axion mass to depend as well on the
temperature of the Universe  
\begin{equation}
 m (T) \simeq 
 \begin{dcases}
    m \left(\frac{\Lambda}{T}\right)^Q
       & \mathrm{for}\, \, \, T > \Lambda , \\
    m 
       & \mathrm{for}\, \, \, T < \Lambda,
 \end{dcases}
\label{eq:mofT}
\end{equation}
with a positive power $Q$ usually of order unity, as understood via non-perturbative estimations in the strongly coupled regimes. As we are only interested in in the $T < \Lambda$ regimes we consider no dependence on T.} 
\begin{equation}
 m_0 = m = \frac{m_u m_d}{m_u+m_d}\frac{\Lambda^2}{f_{\phi}}.
\end{equation}
Without loss of generality we may take the axion to lie within the range
$-\pi \leq a \leq \pi$ with $a=\frac{\phi}{f_{\phi}}$.

%\ag{Call Fig. 1 and write the importance of mu, md ratios.}

\medskip

%\section{Axion-like Particle as the Inflaton}

%\ag{To be filled in with notes from Alessio. And %re-produce ns - r plots.}

%\medskip

\section{ALP as the Curvaton: cosmological evolution}

%{\it Cosmological evolution.}---
%\ag{Explain the temperature dependence and elaborate why temperature dependence is not important}

The vacuum expectation value (VEV) of the curvaton during inflation is always considered to be a free parameter at the classical level here. Nonetheless, in a long-lived inflationary universe, the long
wavelength modes of the quantum fluctuation of a light scalar field may become important, because its Compton wavelength is large compared to the size of the Hubble horizon during inflation. From Ref. \cite{Bunch:1978yq}, we know the vacuum expectation value of the square of such a light scalar field curvaton of mass $m_c$ with potential $V(\sigma_c)=\frac{1}{2}
m_c^2\sigma_c^2$ is given by \be \langle \sigma_c ^2 \rangle={3H_*^4\over 8\pi^2m_c^2}, \ee where $H_*$ is the Hubble parameter during inflation. So the vacuum expectation value of $\sigma_c$ can be estimated
as \be {\sigma_{c}}_*=\sqrt{\langle \sigma_c^2\rangle}=\sqrt{3\over
8\pi^2}{H_*^2\over m_c}. \ee 
Thus for a curvaton field with quadratic
potential, its typical vacuum expectation value is given by
$H_*^2/m$ and its energy density is roughly given by $H_*^4$. This means that larger the
Hubble parameter during inflation, larger the energy density of curvaton. However, the energy density of axion-type curvaton (as in Eqn. \eqref{eq:L}) is bounded by $m^2f^2$ from above. If $\,m f\gg H_*^2$,  $\,{\sigma_c}_*$ can be estimated as $H_*^2/m$. On the other hand, ${\sigma_c}_*\sim f$ if $mf\leq H_*^2$. Since it is difficult to get analytic results for the axion-type curvaton model if $\,{\sigma_c}_*\sim f$, we will use numerical method to calculate the curvature perturbation in the next
section.

Upon analyzing the axion dynamics, we will make a couple of assumptions along the way.

\begin{itemize}
    \item We first assume the hierarchy of energy scales  
$T_{\mathrm{inf}} = H_{\mathrm{inf}} / 2 \pi $.
\begin{equation}
 T_{\mathrm{inf}} < \Lambda < f_{\phi}.
\end{equation}
During inflation the U(1) is already broken.
\item Then we assume $m^2 \ll H_{\mathrm{inf}}^2$, and that
$f_{\phi}$ is smaller than the reduced Planck scale
$M_{\mathrm{Pl}} = (8 \pi G)^{-1/2}$,
then $\Lambda^4 \ll M_{\mathrm{Pl}}^2 H_{\mathrm{inf}}^2$,
i.e., the energy density of the axion is tiny compared to the total
density of the universe 
and thus the axion serves as a spectator field.
\item We assume axion rolling is classical and it rolls slowly and satisfies,

\begin{equation}\label{slowroll}
3 H_{\mathrm{inf}} \dot{\phi} \simeq -\frac{d V(\phi)}{d \phi} .
\end{equation}

For cosine like potential choice $V(a)=\Lambda^4(1-\cos{a})$ integrating this by ignoring any time dependence of the Hubble rate \citep{Kobayashi:2020xhm}
yields 
\begin{equation}
 \tan \left( \frac{a_{\mathrm{end}}}{2}  \right) \simeq
 \tan \left( \frac{a_1}{2}\right) 
\cdot e^{\left( - \frac{m^2}{3 H_{\mathrm{inf}}^2}\mathcal{N}_1
  \right)},
\label{eq:tan_theta}
\end{equation}
with the identification, $m^2=\frac{\Lambda^4}{f_{\phi}^2}$.
Here $a_{\mathrm{end}}$ is the field value when inflation ends,
and $\mathcal{N}_1$ is the number of $e$-folds since the time when the
axion takes a value~$a_1$ until the end of inflation. 
\end{itemize}

Denoting the axion decay rate as $\Gamma_{a}$\footnote{This decay width $\Gamma_a$ is generated due to axion couplings to fermions or gauge bosons or to other scalars, however we do not discuss the microscopic details of the ALP model and instead resort to the decay as a phenomenological parameter. For BSM model details for such $\Gamma_a$ the reader is referred to Ref. \cite{DiLuzio:2020wdo}.},
if $\Gamma_{a} < (>) \, H_{\mathrm{dom}}$, then 
the axion would decay after (before) dominating the universe.
Supposing that the axion decays suddenly when $H = \Gamma_{a}$
(we denote quantities at this time by the subscript ``dec''),
one finds a relation:
\begin{equation}
 3 M_{\mathrm{Pl}}^2 \Gamma_{a} ^2 = 
 \rho_{r,osc}
 \left(\frac{a_{\mathrm{osc}}}{a_{\mathrm{dec}}}\right)^4 
+ \rho_{a,osc}
\left(\frac{a_{\mathrm{osc}}}{a_{\mathrm{dec}}}\right)^3 .
\label{eq:sum}
\end{equation}

%\ag{Add solution details}

The terms in the right hand side correspond to the radiation
and axion densities right before the decay,
and we write their ratio as
$R \equiv (\rho_{a} / \rho_{\mathrm{r}} )_{\mathrm{dec}}$
for later convenience.

\medskip

\section{Curvature perturbation and Non-Gaussianities}

\subsection{Analytical Solutions}
%{\it Curvature perturbation.}---
Converting isocurvature perturbations to adiabatic ones, the conversion is completed  when the axion decays.  Using the $\delta \mathcal{N}$~formalism~\cite{Sasaki:1995aw},
we compute the axion-induced curvature perturbation as the
fluctuation in the number of $e$-folds~$\mathcal{N}$ between an initial
flat slice during inflation when a comoving wave number~$k$ of
interest exists the horizon ($k = a H$),
and a final uniform-density slice where $H = \Gamma_{a} $. Power spectrum,
 and Non-Gaussianities 
(local bi-spectrum and tri-spectrum) 
%\ag{Please insert the expression for gNL here...}
\begin{align}
    P_\zeta (k) \simeq (\partial \mathcal{N} / \partial a_k)^2
(H_k / 2 \pi f_{\phi})^2
\end{align}
\begin{align}
    f_{\mathrm{NL}} \simeq (5/6) 
(\partial^2 \mathcal{N} / \partial a_k^2) 
(\partial \mathcal{N} / \partial a_k)^{-2} 
\end{align}
\begin{align}
    g_{\mathrm{NL}} \simeq (25/54) 
(\partial^3 \mathcal{N} / \partial a_k^3) 
(\partial \mathcal{N} / \partial a_k)^{-3} 
\end{align}
%\ag{Please check these definitions...}
%\ag{Insert the def of $P_r(k)$}

\iffalse
The derivatives of~$\mathcal{N} = \ln (a_{\mathrm{dec}} / a_k)$ can be
evaluated by differentiating (\ref{eq:sum}) multiple times with respect
to $a_k$. 
\footnote{In the vicinity of $a = 0$ where the axion potential is
well-approximated by a quadratic, the mass-to-Hubble ratio~$c$ at the
onset of the oscillation
is independent of the axion field value~\cite{quadratic_c};
however this is not the case in the region close to the hilltop
$\lvert a \rvert = \pi$ \cite{Kawasaki:2011pd}.}
Let us for the moment suppose that 
$a_{\mathrm{end}}$ is not too far from $0$
and take $\partial c / \partial a_{k}  = 0$.
Then, also because the axion density is negligibly tiny 
compared to the total density of the universe during the scale
factor
$a_k \leq a_i \leq a_{\mathrm{osc}}$,
the ratio $(a_{\mathrm{osc}} / a_k)$ is independent of
$a_k$. 
Derivatives of $a_{\mathrm{end}}$ can be evaluated
using~(\ref{eq:tan_theta}).  
Hence, after some manipulation we find 
\fi

Now if we consider a cosine potential for the axion field $V(a) = \Lambda^4 \left[1- \cos (a)\right]$ the 
above power spectrum and bispectrum can be written as \cite{Kawasaki:2011pd,Kobayashi:2020xhm}:
%\ag{Edit here...}
\begin{widetext}
\begin{equation}
\begin{split}
& P_\zeta (k) \simeq
\left( \frac{R}{3 R + 4} 
\frac{1 + \cos a_{\mathrm{end}}}{\sin a_k}\frac{H_k}{2 \pi f_{\phi}}
\right)^2 
%\quad
\\
& f_{\mathrm{NL}} \simeq
\frac{5}{6} \left\{
\frac{4}{R} + \frac{4}{3 R + 4}
-  \left( 3 + \frac{4}{R} \right)
\frac{1 - \cos a_{\mathrm{end}} + \cos a_k}{1 + \cos
a_{\mathrm{end}}} 
\right\},
\\
\end{split}
\label{eq:Pzeta-fNL}
\end{equation}

and the tri-spectrum can be written as,
\begin{multline}
    g_{\mathrm{NL}} \simeq 
\frac{25}{54} \left\{ -4 (1+R)\left(\frac{4}{R^2}+\frac{12}{(3R+4)^2}\right)+ 2
\left(\frac{4}{R}+\frac{4}{(3R+4)}\right)^2 \right.  \\
\left. - 3 \left(3+\frac{4}{R}\right)\left(\frac{4}{R}+\frac{4}{3R+4}\right)\frac{1-\cos{a_{end}}+\cos{a_k}}{1+\cos{a_{end}}}  + \left(3+\frac{4}{R}  \right)^2 \right. \\ 
\left. \left[\left(\frac{1-\cos{a_{end}}+\cos{a_k}}{1+\cos{a_{end}}}\right)^2  +
\frac{\cos^2{a_{end}}+\cos{a_k}-\cos{a_k}\cos{a_{end}}}{(1+\cos{a_{end}})^2} \right] \right\}
\label{eq:gNL}
\end{multline}
\end{widetext}

The spectral index of the power spectrum 
$n_s - 1 = d \ln P_\zeta / d \ln k$ 
can be computed using $d \ln k \simeq H_k d t$ and the slow-roll
approximation during inflation as
\begin{equation}
 n_s - 1 \simeq 
\frac{2}{3} \frac{V^{''}(\phi) }{H_k^2} \cos a_k
+ 2  \frac{\dot{H}_k}{H_k^2}.
\label{eq:n_s-1}
\end{equation}

We would work in $R \rightarrow \infty$ and it is easy to see that when $a \rightarrow 0$ this 
expressions produce the correlation functions of vanilla axion scenario. In general the last term in (17) is the first
slow roll parameter and is tightly constrained from tensor-to-scalar ratio or equivalently from
the single field consistency relation. But in the case of curvaton scenario as the curvaton field
is sourcing the perturbations the last term in (17) can be larger than the Planck constraint
and can be estimated as 0.02 \cite{Gordon:2002gv}.

\begin{figure}
    \centering
      \includegraphics[width=0.6\textwidth]{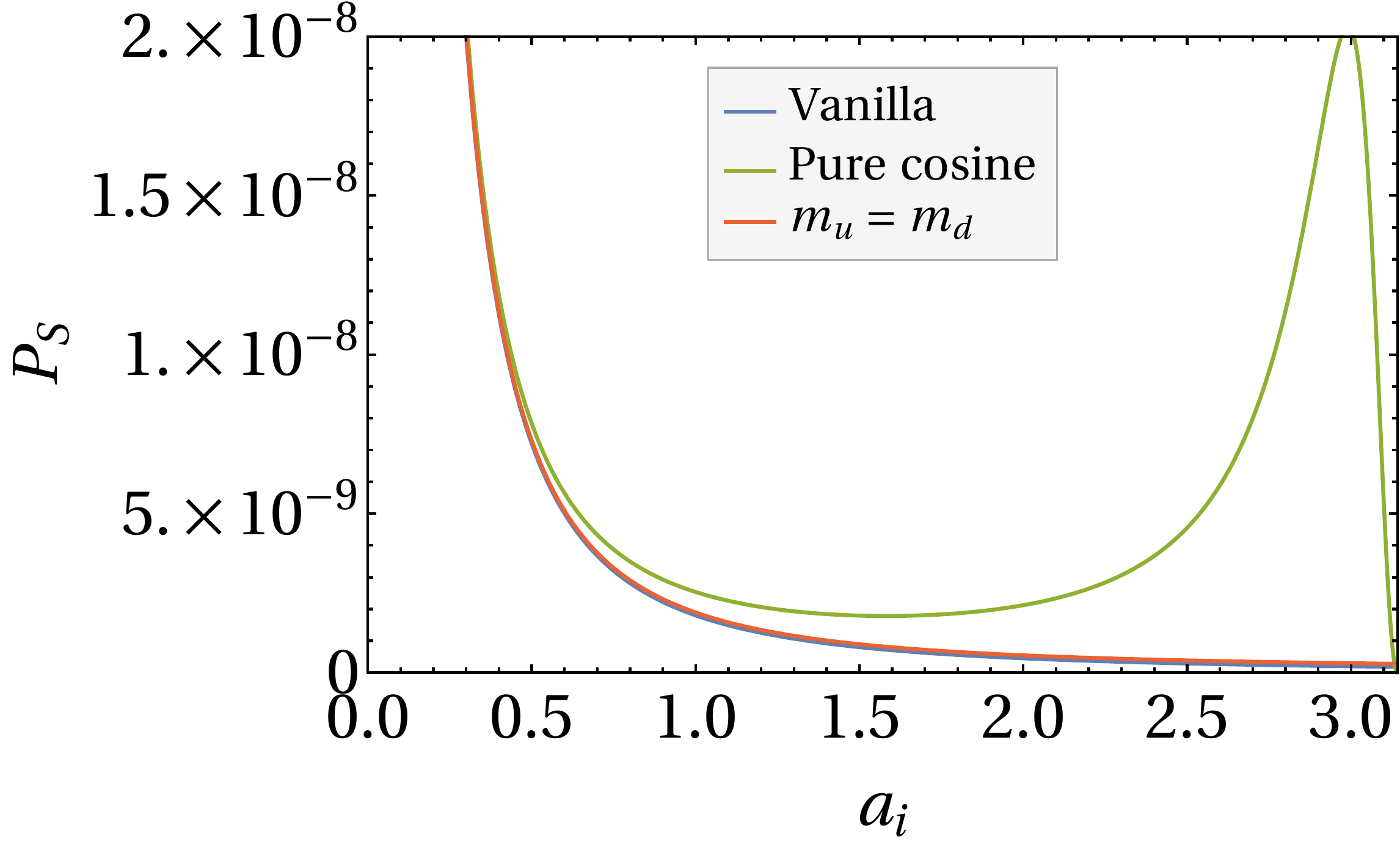}    
        \includegraphics[width=0.49\textwidth]{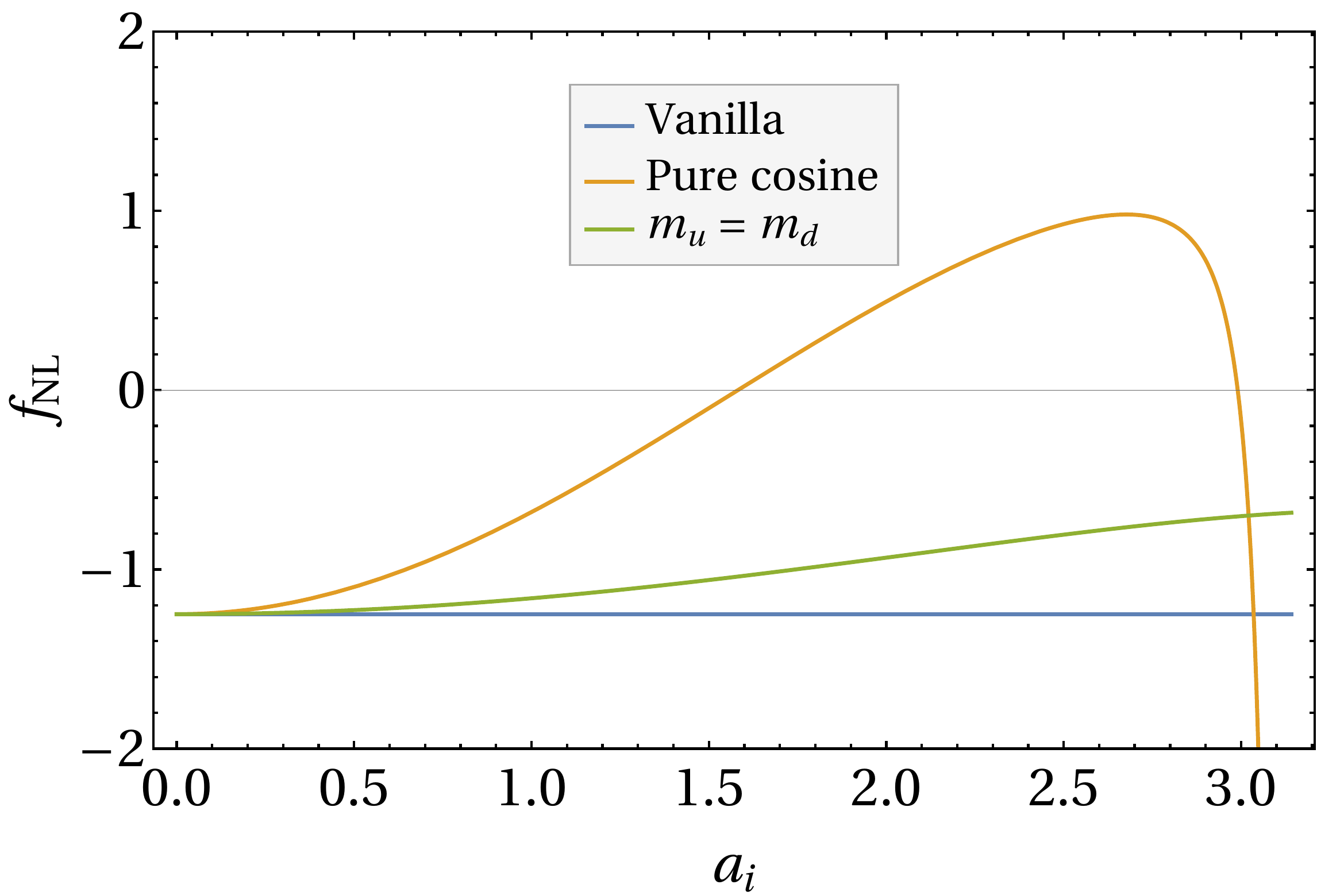}   
        \includegraphics[width=0.48\textwidth]{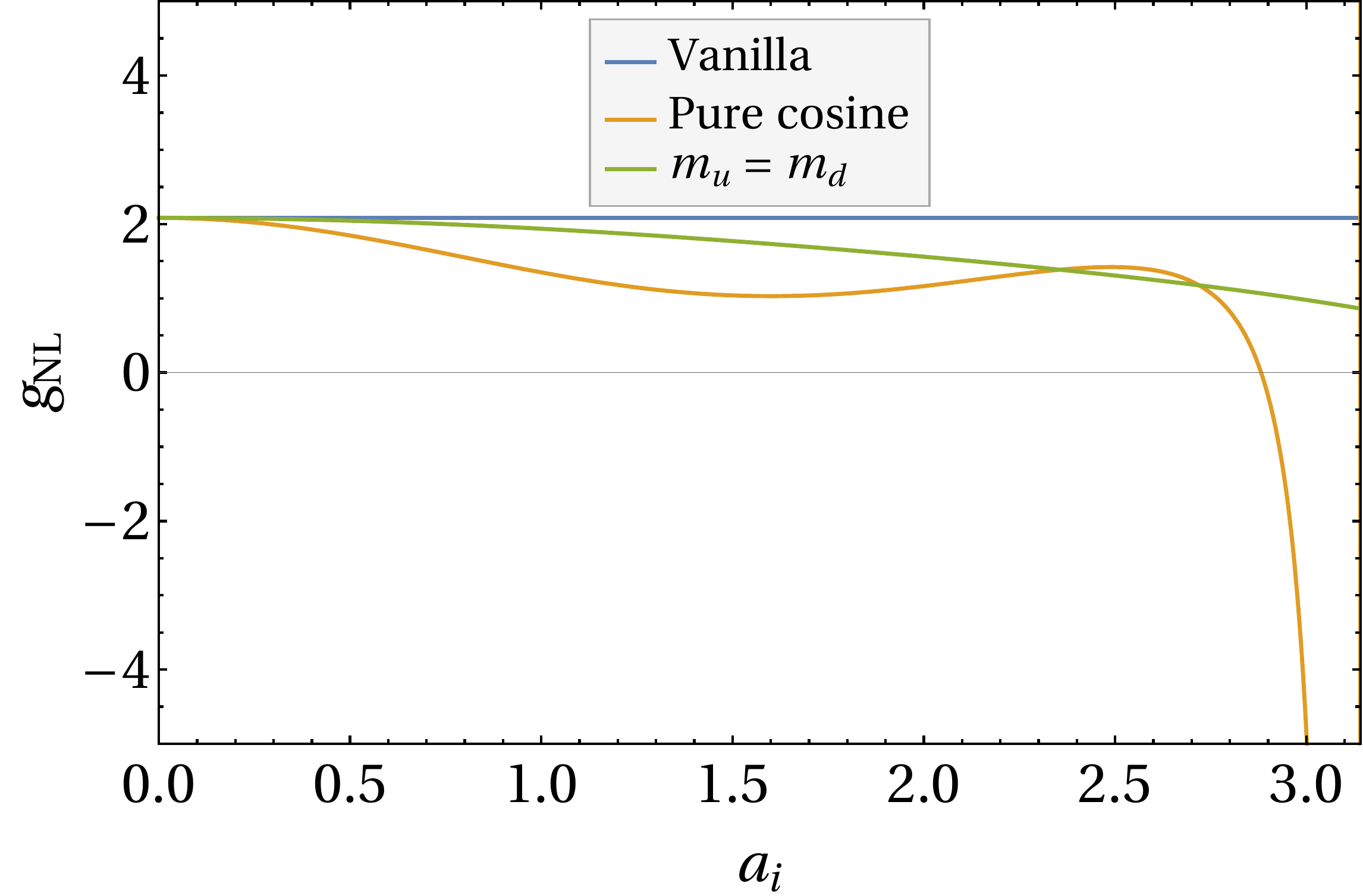}    
    \caption{\it Analytical estimates for the scalar power spectrum, bi-spectrum and 
    tri-spectrum. It can be noted that for $m_u=m_d$ limit the power spectrum
    behaves like vanilla scenario. The correlation functions for cosine potential has distinct nature
    compared to the vanilla case and $m_u=m_d$ case as in this case the correlation 
    functions depend non-trivially on $a_i$. Also for $m_u=m_d$ case the bispectrum and trispectrum
    both depend on $a_i$ unlike the vanilla scenario. Parameter choice for all three plots are 
    $H_{\rm inf}=10^{-5} M_{\mathrm{Pl}}, f_{\phi} = 2500 H_{\rm inf}$ \& $\Lambda=30 H_{\rm inf}$.}
    \label{fig:analytical}
\end{figure}
%\ag{Are the fNL and gNl expressions are same for both the limits ?}.

One important thing to note here is that the nature of potential \eqref{main-pot} becomes 
$V = \Lambda^4 (1-\cos (a/2))$ in $m_u=m_d$ limit (see \eqref{pot:mu=md})
and from here we can easily estimate the 
power spectrum for Eqn. \eqref{main-pot} potential in $m_u=m_d$ limit as,

\begin{equation}
    P_\zeta (k) \simeq
\left( \frac{R}{3 R + 4} 
\frac{1 + \cos a_{\mathrm{end}}}{\sin a_k}\frac{H_k}{2 \pi (2f_{\phi})}
\right)^2 ~~~  \text{for}~~~  m_u = m_d.
\label{analytical-Ps}
\end{equation}

and the higher order correlation functions for this scenario can be written as,

\begin{equation}
    f_{\mathrm{NL}} \simeq
\frac{5}{6} \left\{
\frac{4}{R} + \frac{4}{3 R + 4}
-  \left( 3 + \frac{4}{R} \right)
\frac{1 - \cos \frac{a_{\mathrm{end}}}{{2}} + \cos \frac{a_k}{2}}{1 + \cos
\frac{a_{\mathrm{end}}}{2}} 
\right\}
\label{analytical-fNL}
\end{equation}

\begin{multline}
    g_{\mathrm{NL}} \simeq 
\frac{25}{54} \left\{ -4 (1+R)\left(\frac{4}{R^2}+\frac{12}{(3R+4)^2}\right)+ 2
\left(\frac{4}{R}+\frac{4}{(3R+4)}\right)^2 \right.  \\
\left. - 3 \left(3+\frac{4}{R}\right)\left(\frac{4}{R}+\frac{4}{3R+4}\right)\frac{1-\cos{\frac{a_{end}}{2}}
+\cos{\frac{a_k}{2}}}{1+\cos{\frac{a_{end}}{2}}}  + \left(3+\frac{4}{R}  \right)^2 \right. \\ 
\left. \left[\left(\frac{1-\cos{\frac{a_{end}}{2}}+\cos{\frac{a_k}{2}}}{1+\cos{\frac{a_{end}}{2}}}\right)^2  +
\frac{\cos^2{\frac{a_{end}}{2}}+\cos{\frac{a_k}{2}}-\cos{\frac{a_k}{2}}\cos{\frac{a_{end}}{2}}}{(1+
\cos{\frac{a_{end}}{2}})^2} \right] \right\}
\label{analytical-gNL}
\end{multline}

\medskip

From the analytical expressions for correlation functions as shown in Fig \ref{fig:analytical} and we can see that the nature of power spectrum for vanilla axion case and $m_u=m_d$ limit of potential \eqref{main-pot}
are almost identical as it is evident from the nature of \eqref{main-pot} which suggests that as $m_u$ becomes
equal to $m_d$ the potential behaves more like a vanilla axion scenario (see fig \ref{fig:potential}).
The nature of \eqref{main-pot} also suggests that as $m_u$ becomes larger than $m_d$ the potential starts 
to behave more as a pure cosine kind of axion potential and we can expect that
for $m_u \gg m_d$ limit the power spectrum will behave closely as Eqn. \eqref{eq:Pzeta-fNL}. The power spectrum in the pure cosine case has a distinct behavior compared to the vanilla case and $m_u = m_d$ case. The bispectrum for pure cosine limit exhibits positive and negative $f_{NL}$ values depending on the value of $a_i$. The trispectrum becomes negative at large $a_i$ value. It is also
important to note here that the nature of bispectrum and trispectrum for $m_u=m_d$ case is completely
different than vanilla scenario and these higher order correlation functions have $a_i$ dependence as 
opposed to vanilla case where these higher order correlation functions are perfectly constant. This is expected as for $m_u=m_d$ case as the potential does not exactly boil down to vanilla scenario, but the power spectrum can not distinguish between these natures while bispectrum and trispectrum can. Also, the bispectrum remains negative for all $a_i$ values and trispectrum remains positive for all $a_i$ values for both the cases.

\begin{figure}
    \centering
      \includegraphics[width=0.6\textwidth]{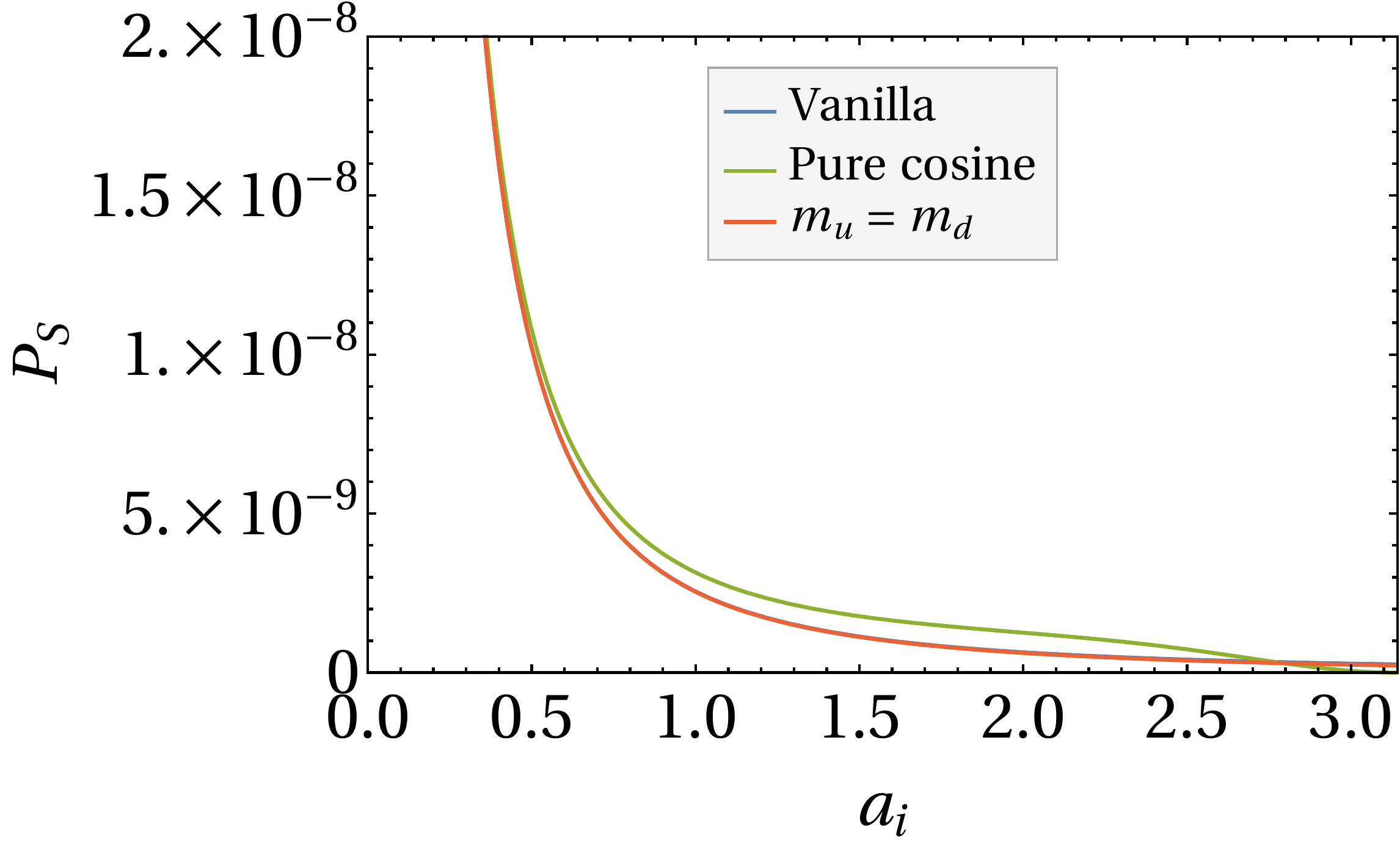}    
        \includegraphics[width=0.49\textwidth]{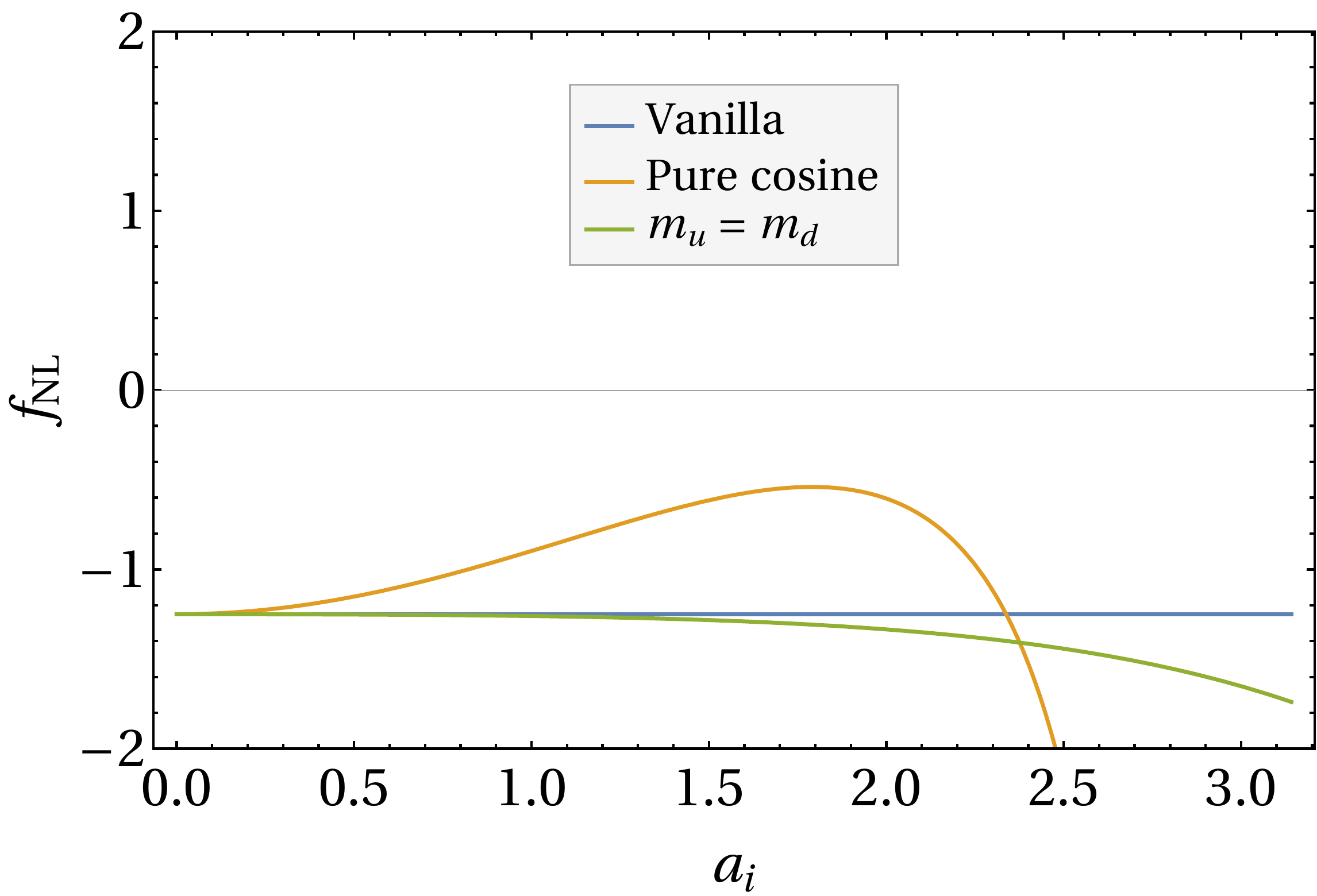}   
        \includegraphics[width=0.48\textwidth]{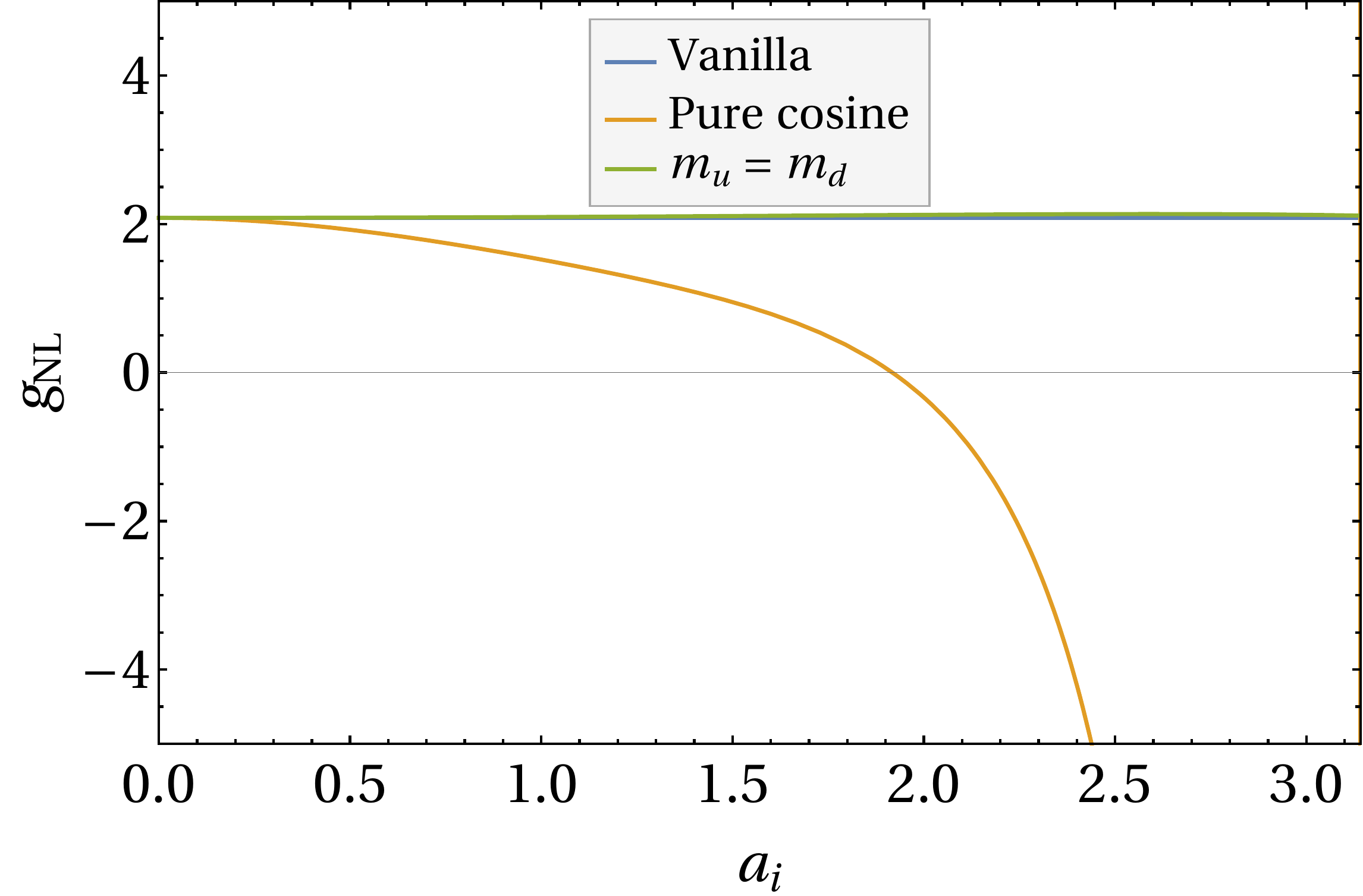}    
    \caption{\it Analytical estimates for the scalar power spectrum, bi-spectrum and 
    tri-spectrum for
    $H_{\rm inf}=10^{-5} M_{\mathrm{Pl}}, f_{\phi} = 2100 H_{\rm inf}$ \& $\Lambda=20 H_{\rm inf}$.}
    \label{fig:analytical2}
\end{figure}

In Fig. \ref{fig:analytical2} we have plotted the power spectrum, bispectrum and trispectrum for $m_u = m_d$, pure axion and vanilla scenarios for $f_{\phi} = 2100 H_{\rm inf}$ and $\Lambda=20 H_{\rm inf}$. Comparing with Fig. \ref{fig:analytical} we can see that the behavior of the correlation functions completely changes for this choice of the parameters of the model. Here the power spectrum for pure axion case mostly follows the vanilla-like behavior with a larger amplitude for large $a_i$ values. But the bispectrum has only negative values for all $a_i$ values even for the pure axion case. Also, the trispectrum becomes negative for a smaller $a_i$ value than the previous case for the pure axion case and while it remains constant for both vanilla and $m_u=m_d$ cases.

Though the correlation functions for $m_u=m_d$ and pure axion case  can be derived exactly analytically it becomes complicated to evaluate the correlation function in different limits of $m_u$ and $m_d$ analytically, so in the next section we will discuss the numerical evaluation of the correlation functions. 

\subsection{Numerical Solution}

In this section, we describe the detailed analysis of the results obtained. So far we have introduced the analytical expressions for the scalar power spectrum, bi-spectrum and tri-spectrum. In order to complete our analysis we would focus on numerical solutions also. The evolution
of the axion field during inflation will be governed by \cite{Dimopoulos:2003az,Chingangbam:2009xi}:

\begin{equation}
    3 H_{\mathrm{inf}} \dot{a} \simeq -\frac{d V(a)}{d a}. 
\end{equation}

For the post-inflationary period we have solved the full set of Friedmann equations along with Klein-Gordon equation for axion field a, 

\begin{equation}
\begin{split}
&  H_{inf}^2 = \frac{1}{3 M_{\mathrm{Pl}}^2} \left(\rho_r + \rho_{a}\right)
\\
& \dot{\rho}_r + 4 H \rho_r = 0
\\ 
& \rho_{a} = \frac{1}{2} \dot{a}^2 + V(a)
\\
& \ddot{a} + 3 H \dot{a} + \frac{d V(a)}{d a} = 0
\end{split}
\end{equation}

%\ag{We need to change the notations....from $\phi$ to a.}
%\ag{Please define all the quantitites above.} 
Here $\rho_r$
and $\rho_{a}$ are energy densities of radiation and axion field
respectively.
These four set of equations can be converted into two set of coupled equations with redefined 
variables, $x = m t$ as \cite{Chingangbam:2009xi},

\begin{equation}
    \begin{split}
        & \frac{d N}{d x} = \left[\alpha e^{-4 N}+\frac{f_{\phi}^2}{3 M_{\mathrm{Pl}}^2} \left\{
        \frac{1}{2}\left(\frac{d a}{d x}\right)^2 + V(a)\right\}\right]^{\frac{1}{2}}
        \\
        & \frac{d^2 a}{d x^2} = - 3 \frac{d N}{d x} \frac{d a}{d x} - \frac{d V(a)}{d a}
    \end{split}
\end{equation}

Here, $\alpha = \rho_{r,0}/(3 m^2 M_{\mathrm{Pl}}^2)$ and can be considered $\mathcal{O}(1)$
because at the end of inflation the energy density of radiation was dominant and so at
$H_{inf}=m$, $\rho_{r,0} = 3 m^2 M_{\mathrm{Pl}}^2$.
The numerical integration would be done from the time when the axion starts to 
oscillate at $H_{inf}=m$ i,e $x=1$ until the axion field decays at $H_{inf}=\Gamma_{a}$ i,e 
$x=\frac{m}{\Gamma_{a}}$. One important point to notice here is that in the equation of motion of the axion field we do not consider the effect of the decay width 
$\Gamma_{a}$ as we are working in $R \rightarrow \infty$ and  as a result $\Gamma_{a}$ is 
very small compared to the axion mass and here we are considering
$\Gamma_{a} / m \sim 10^{-16}$.

%\ag{Please describe why are we taking this limit, and its connection to delta-N formalism..}

\medskip

\section{Matching with PLANCK Observations}

In this section, we compute the power spectrum, bi-spectrum and tri-spectrum numerically and normalize with the Planck data.

\begin{figure}[H]
    \centering
      \includegraphics[width=0.6\textwidth]{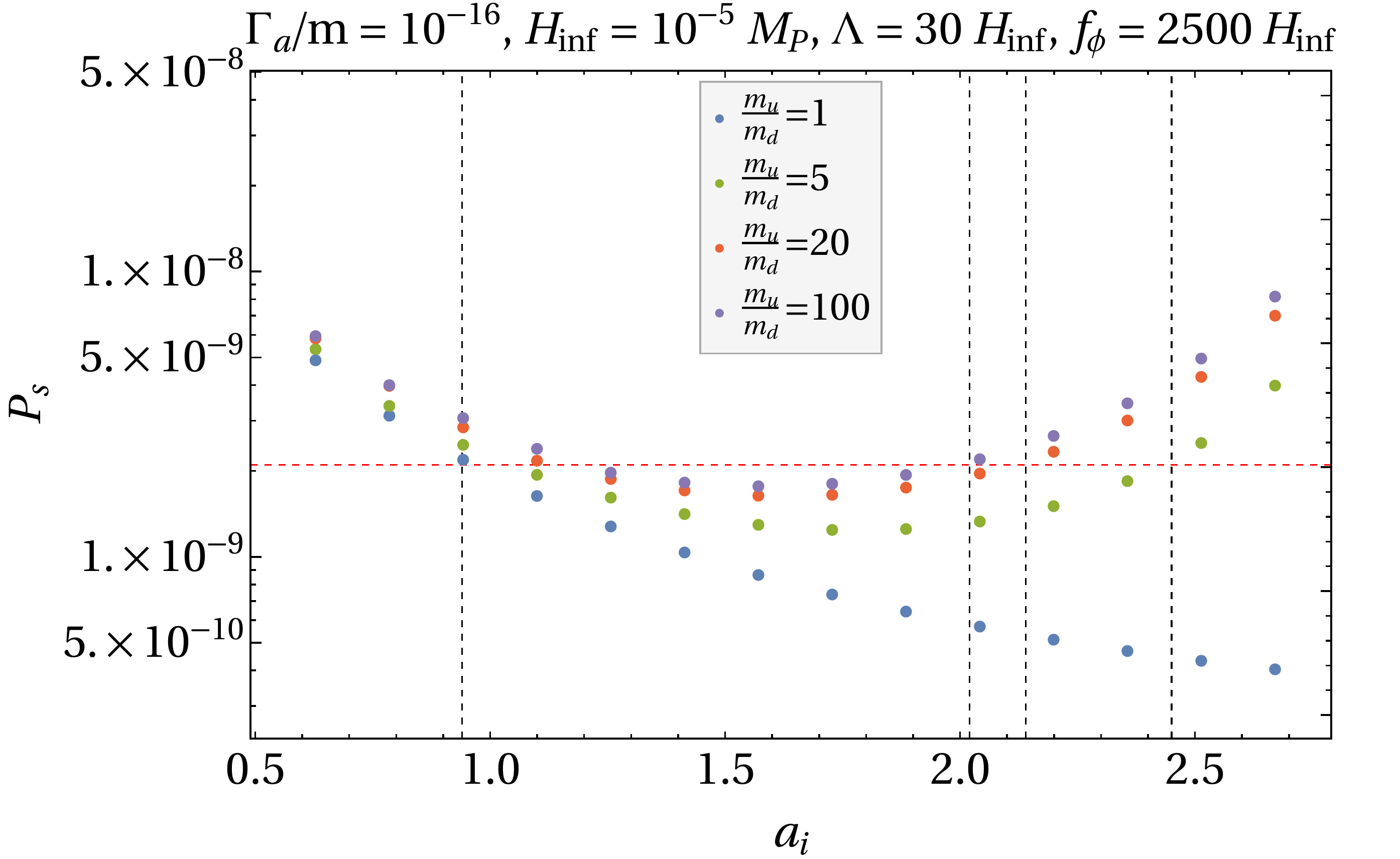}    
        \includegraphics[width=0.48\textwidth]{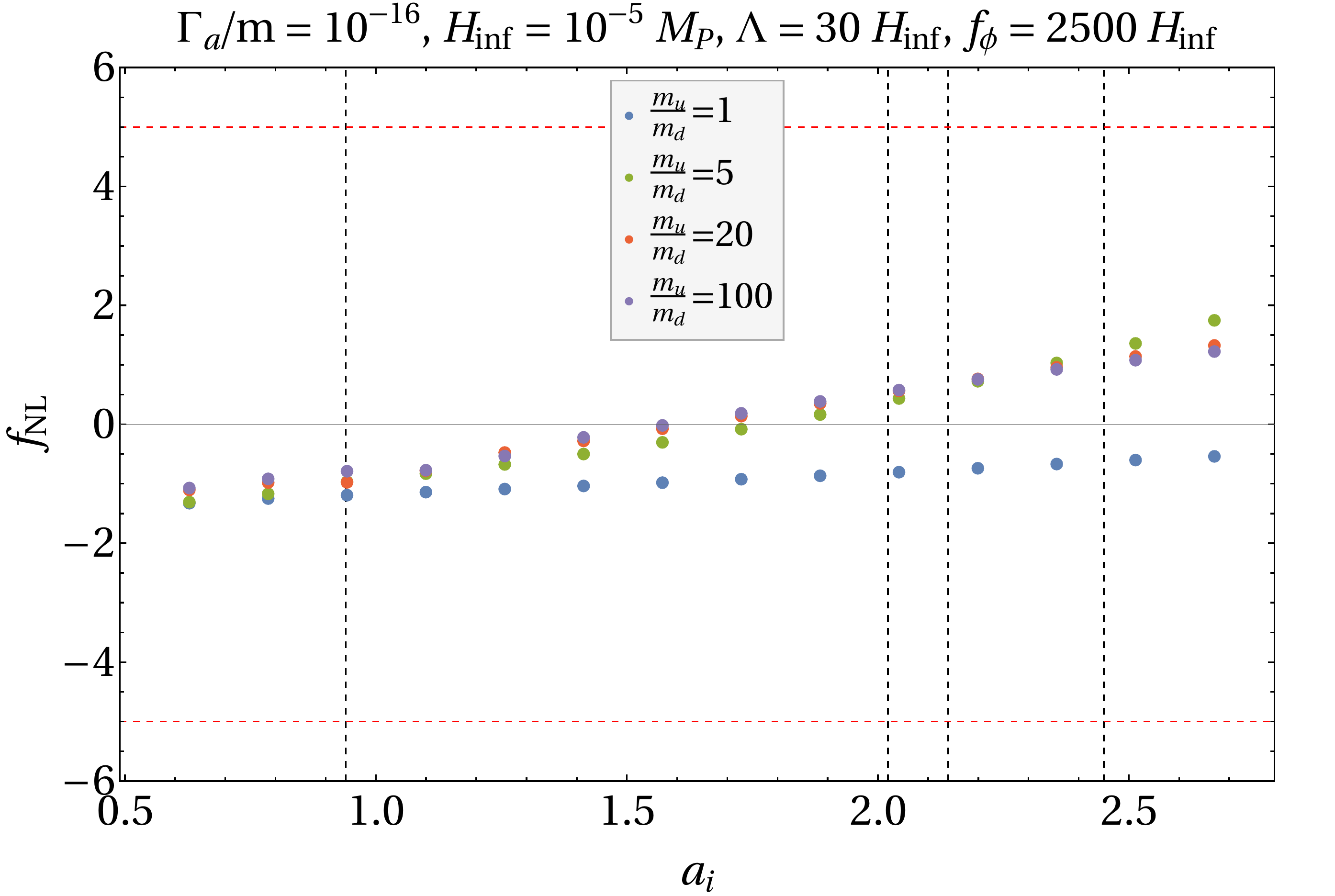}   
        \includegraphics[width=0.48\textwidth]{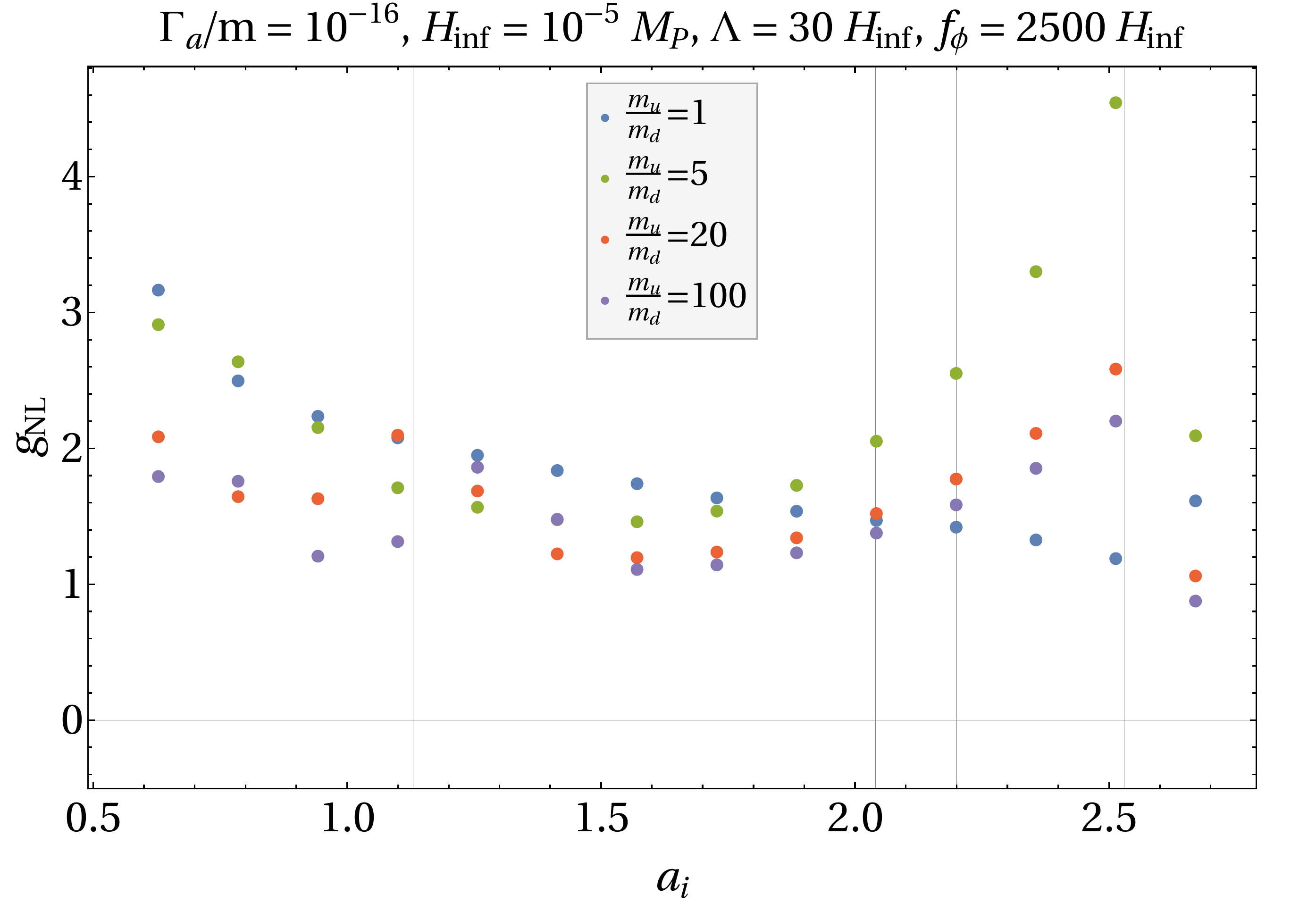}    
    \caption{\it Plots of scalar power spectrum $P_s$, scalar bi-spectrum $f_{NL}$ and scalar tri-spectrum $g_{NL}$ versus $a_i$. We show the dependence of the spectra on various combinations of $m_u$ and $m_d$ masses for the parameters $\frac{\Gamma_{a}}{m}=10^{-16}$, $H_{\rm inf}=10^{-5} M_{\mathrm{Pl}}, f_{\phi} = 2500 H_{\rm inf}$ \& $\Lambda=30 H_{\rm inf}$. The horizontal red lines show Planck's bound on the respective observables.
    The vertical dotted lines describe the value of $a_i$ where the Planck normalization for the scalar power spectrum is satisfied for different $m_u$ and $m_d$ limits. Here we only include
    the $a_i$ range where we can trust our numerical results.
}
    \label{fig:numerical1}
\end{figure}
%\subsection{The Constraint from Power Spectrum amplitude}

\begin{figure}[H]
    \centering
      \includegraphics[width=0.6\textwidth]{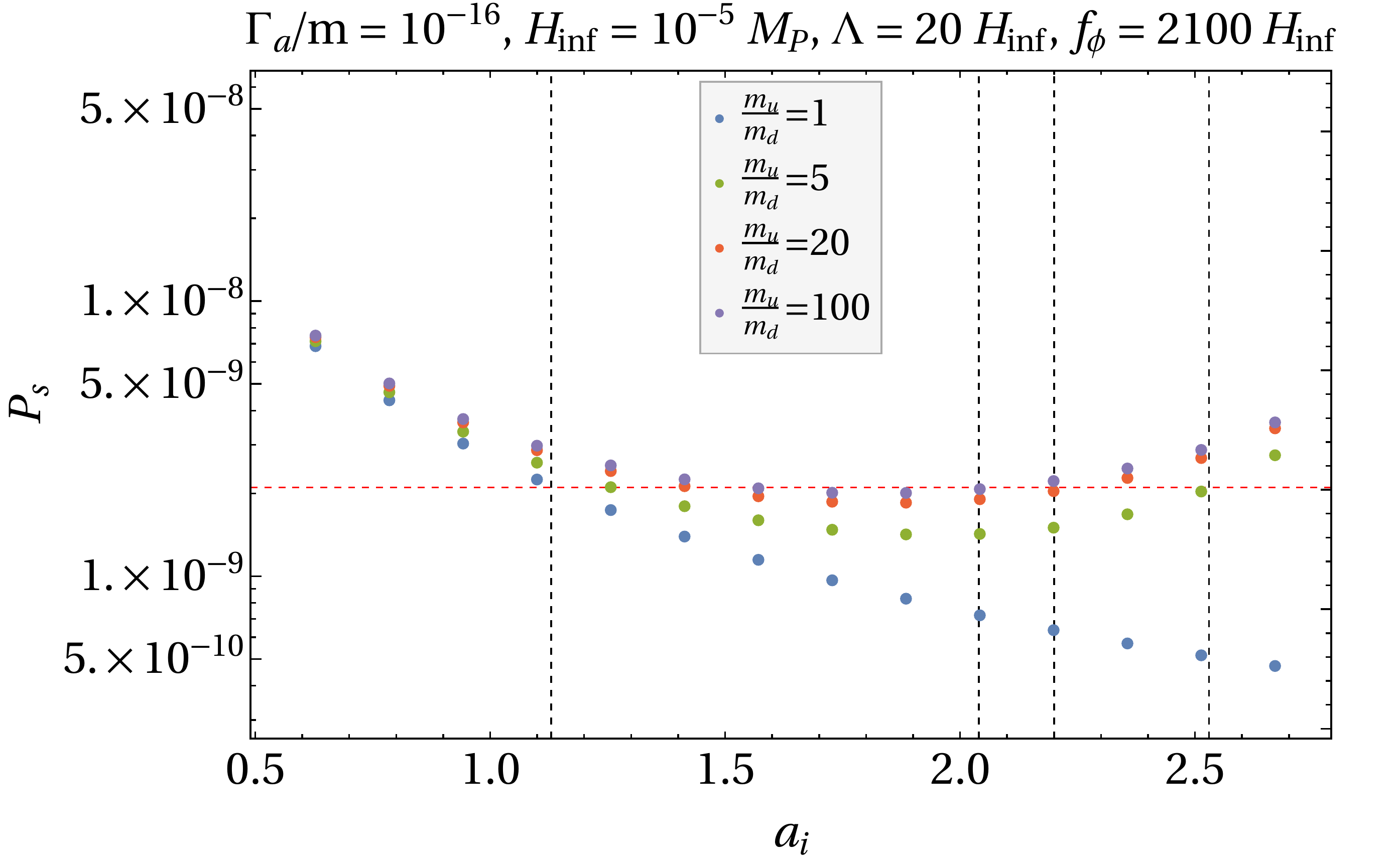}    
        \includegraphics[width=0.48\textwidth]{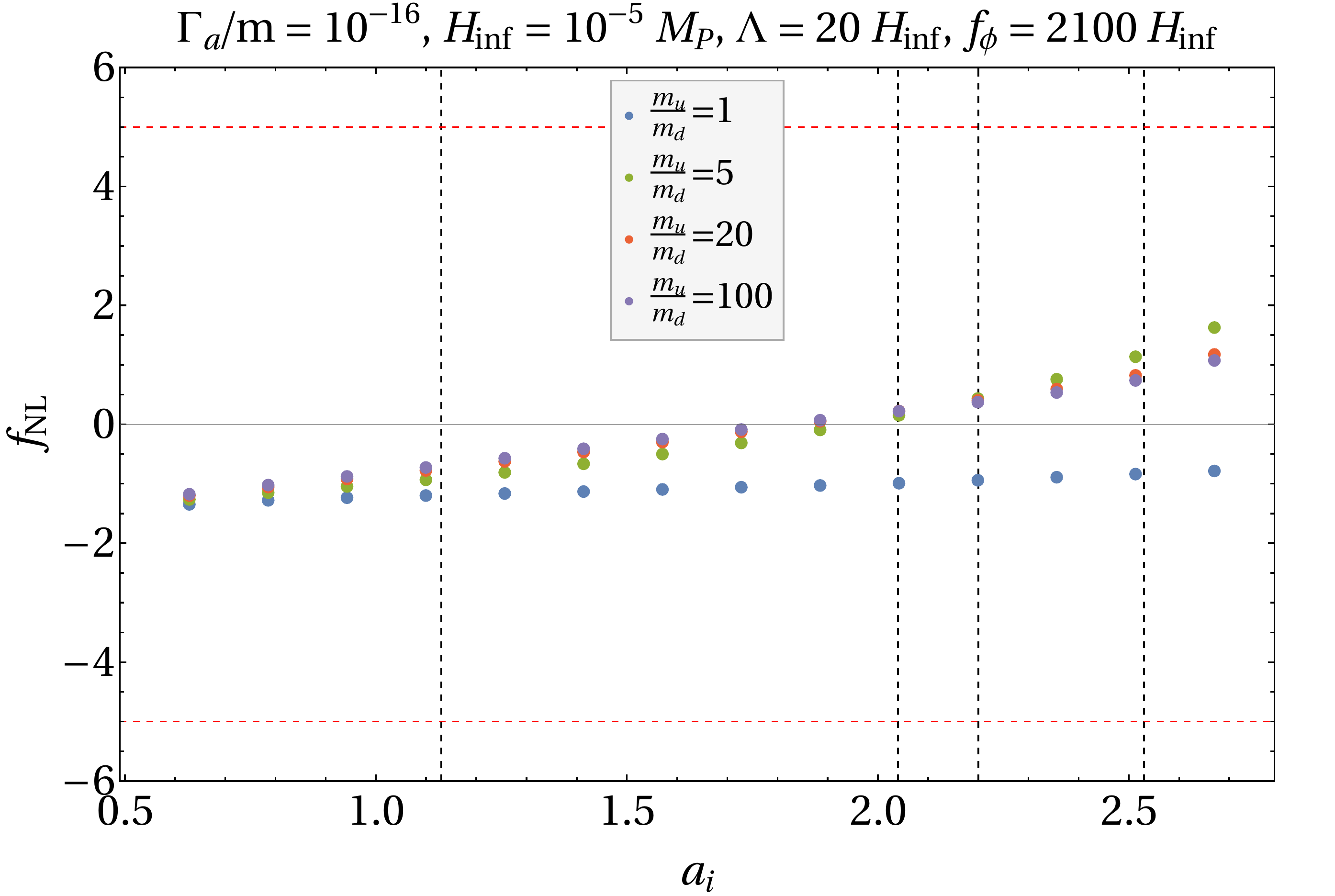}   
        \includegraphics[width=0.48\textwidth]{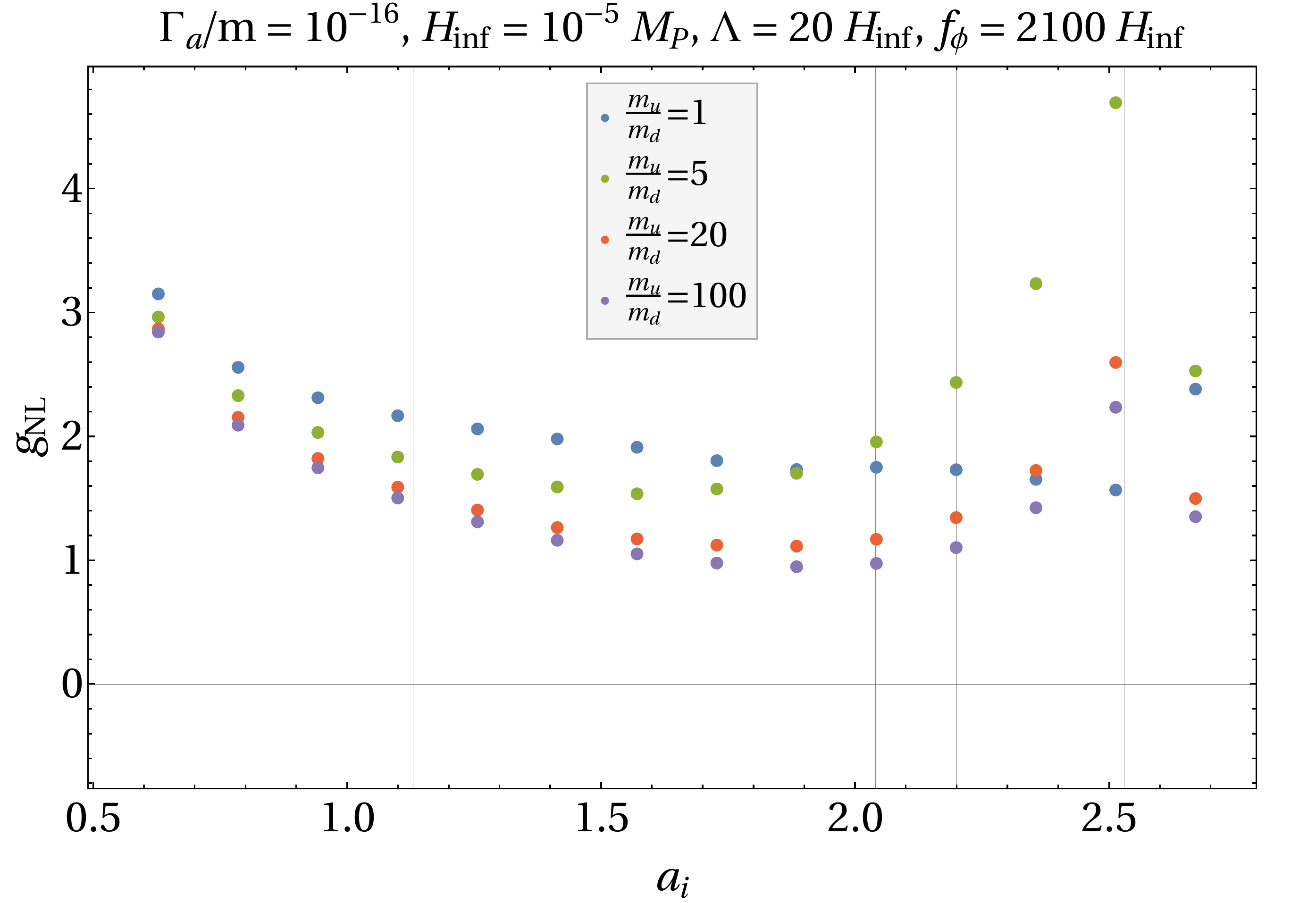}    
    \caption{\it Plots of scalar power spectrum $P_s$, scalar bi-spectrum $f_{NL}$ and scalar tri-spectrum $g_{NL}$ versus $a_i$. We show the dependence of the spectra on various combinations of $m_u$ and $m_d$ masses for the parameters $\frac{\Gamma_{a}}{m}=10^{-16}$, $H_{\rm inf}=10^{-5} M_{\mathrm{Pl}}, f_{\phi} = 2100 H_{\rm inf}$ \& $\Lambda=20 H_{\rm inf}$. The horizontal red lines show Planck's bound on the respective observables.
    The vertical dotted lines describe the value of $a_i$ where the Planck normalization for the scalar power spectrum is satisfied for different $m_u$ and $m_d$ limits. Here we only include
    the $a_i$ range where we can trust our numerical results.
}
    \label{fig:numerical2}
\end{figure}

\iffalse
\begin{figure}[H]
    \centering
      \includegraphics[width=0.6\textwidth]{plots/Set2-Ps.pdf}    
        %\includegraphics[width=0.47\textwidth]{Pr.pdf}  
        \includegraphics[width=0.48\textwidth]{plots/Set2-fNL.pdf}   
        \includegraphics[width=0.48\textwidth]{plots/Set2-gNL.pdf}    
    \caption{\it \ag{Change the labels and cut out controversial regions.} Plots of scalar power spectrum $P_s$, tensor power spectrum Pr, scalar bi-spectrum $f_{NL}$ and scalar tri-spectrum $g_{NL}$ versus $a_i$. We show the dependence of the spectra on various combinations of $m_u$ and $m_d$ masses for the parameters $H=10^{-5}M_{\mathrm{Pl}}, ~ \Lambda=30 H, f_{\phi}=2400 H, \frac{\Gamma_{a}}{m}=10^{-16}$. The horizontal red lines show Planck's bound 
    on the respective observables.
    The vertical dashed lines describe the value of $a_i$ where the Planck normalization for the scalar power spectrum is satisfied for different $m_u$ and $m_d$ limits. The difference in different $m_u$ and $m_d$ limits is mostly visible in the power spectrum where in $m_u=m_d$ limit the axion behaves as vanilla axion scenario. But in $m_u > m_d$ limit the axion behaves as the cosine axion model. 
    %\ag{Vanilla curvaton nature is recovered for.....}. \ag{mu/md variation is not visible in the plots..}
    }
    \label{fig:numerical2}
\end{figure}
\fi

%\ag{Re-scaling of the axis}
In Figs. \ref{fig:numerical1} and \ref{fig:numerical2} we have analyzed the power spectrum, bi-spectrum and tri-spectrum for the parameter choice: 
$\Gamma_{a}/m = 10^{-16},~ H_{\rm inf} = 10^{-5} M_{\mathrm{Pl}}$, with $\Lambda = 30 H_{\rm inf} , f_{\phi} = 2500 H_{\rm inf}$ and $\Lambda = 20 H_{\rm inf} , f_{\phi} = 2100 H_{\rm inf}$ respectively. We have compared them with recent Planck constraints\footnote{We have included only the range of $a_i$ in Fig. \ref{fig:numerical1} where the numerical result can be trusted.}. If we compare the behavior of the correlators estimated numerically in Fig. \ref{fig:numerical1} with the analytical one in Fig. \ref{fig:analytical} where $\Lambda = 30 H_{\rm inf}, f_{\phi} = 2500 H_{\rm inf}$ we can see that the nature of the numerical estimation closely follows the analytical behavior except for large $a_i$ range. This is because the dependence of the mass-to-Hubble ratio is always taken to be constant and this approximation can break down for large $a_i$ limit \cite{Kobayashi:2020xhm}. But if we compare the numerical estimation of Fig. \ref{fig:numerical2} with Fig. \ref{fig:analytical2} where $\Lambda = 20 H_{\rm inf}, f_{\phi} = 2100 H_{\rm inf}$ we can see that there is a huge disagreement between the analytical and numerical estimation. In the former, we can see that the power spectrum for pure cosine case closely follows the vanilla scenario but in the numerical estimation for $m_u/m_d = 100$ which closely behaves like a cosine potential, we can see that the power spectrum behaves differently compared to the vanilla case and also compared to the $m_u/m_d=1$ case which closely resembles the vanilla scenario. Also in the $m_u/m_d = 100$ limit with numerical estimation, we can see that $f_{NL}$ can have both positive and negative values depending on $a_i$ but in the analytical case, we have seen that with this parameter choice $f_{NL}$ was always negative for pure axion case. This discrepancy arises because in the analytical estimation, we have started with Eqn. \eqref{eq:sum} where we have considered the energy of the axion redshifts as $a^{-3}$. With this consideration, we have performed derivatives of the number of e-folds with respect to the field value to compute the correlation functions. But this approximation breaks down if the potential cannot be approximated as $m^2 a^2$ type. As the axion potential of Eqn. \eqref{main-pot} is different from $m^2 a^2$ type, the analytical and numerical results do not match for most of the parameter space under consideration.

As for the numerical behavior of the correlators, the first thing to notice here is that the scalar power spectrum matches the Planck
bound at different $a_i$ values for different $m_u/m_d$ choices. With  $f_{\phi} = 2500 H_{\rm inf}$ and $\Lambda = 30 H_{\rm inf}$ when $m_u / m_d = 1$ the
power spectrum is normalized at $a_i \sim 0.94$, when $m_u/m_d =5$ at  $a_i \sim 2.45$,
at $a_i \sim 2.14$ for $m_u/m_d =20$ and $a_i \sim 2.02$ when $m_u/m_d =100$.  This behavior can 
be explained from Fig \ref{fig:potential} where we have seen that when $m_u = m_d$
the potential actually behaves very close to as a quadratic potential and the 
nature of the power spectrum is very close to the vanilla case as can also be seen in
Fig \ref{fig:analytical}. Now as $m_u$ starts to get larger than $m_d$ the potential starts to 
behave more as a cosine kind of potential and the nature of the power spectrum for three $m_u\neq m_d$ cases 
discussed above starts to behave as the power spectrum in the case of cosine potential. Although 
one important thing to note here is that the Planck normalization of the power spectrum for these three cases are met at three different values of $a_i$. For same choice of $\Gamma_{a}/m$, $H_{inf}$, $\Lambda$ and $f_{\phi}$ the power spectrum matches the Planck normalization at largest $a_i$ value for $m_u/m_d=5$ case and smallest $a_i$ value for $m_u/m_d=100$ case. So as $m_u$ gets larger than $m_d$ i,e as the potential 
starts to behave more and more like a cosine potential and smaller value of $a_i$ satisfies Planck normalization.

Now if we focus on the bispectrum we can see that the $m_u=m_d$ case has very small $a_i$ dependence
but certainly not constant and $f_{NL} \sim -1.15$ consistent with Planck normalization of the power spectrum. 
Now the bispectrum in the vanilla case is constant $f_{NL} \sim -1.25$; so even if the 
power spectrum for $m_u=m_d$ case and vanilla scenario are very similar in nature there is a clear
distinction in the behavior of bispectrum. For the three $m_u \neq m_d$ cases we see that the largest 
$f_{NL} \sim 1.19$ one can get from $m_u/m_d =5$ and the value for $f_{NL}$ decreases as $m_u$ gets larger
and larger than $m_d$ with $f_{NL} \sim 0.69$ for $m_u/m_d =20$ and $f_{NL} \sim 0.55$ for $m_u/m_d =100$ cases. So in general we can see that as $m_u/m_d$ becomes larger i.e., as the potential becomes more like a cosine potential than the vanilla kind of nature the 
amplitude of bispectrum decreases. The nature of the trispectrum follows the behavior of the bispectrum very closely. Again $m_u/m_d=1$ case
have a $a_i$ dependence which is not observed in the vanilla case and the amplitude of the trispectrum decreases with the increasing value of $m_u/m_d$. 
If we choose $f_{\phi} = 2100 H_{\rm inf}$ and $\Lambda = 20 H_{\rm inf}$ the power spectrum gets normalized at $a_i \sim 0.94$ for $m_u/m_d = 1$, at $a_i \sim 2.52$ for $m_u/m_d = 5$, $a_i \sim 2.19$ for $m_u/m_d = 20$ and $a_i \sim 2.04$ for $m_u/m_d = 100$. So the normalization shifts towards the right in $a_i$ values for this parameter choice compared to the previous case.

One of the most important aspects of our model is that it produces different aspects of ALP behavior in different $m_u/m_d$ limits and upcoming more precise data can help us put bound on the $m_u/m_d$ limit and hence pinpoint the axion behavior. In our analysis, we have seen that the power spectrum can be the main distinguishing factor between the different limits of $m_u/m_d$ as it satisfies the Planck bound in different $a_i$ limits. Also, the estimation of bi-spectrum and tri-spectrum can be a crucial factor as we have seen that the different cases can produce very different $f_{NL}$ and $g_{NL}$ values. Although the scope of exploring the details of the parameter space involving $f_{\phi}$ and $\Lambda$ is beyond the scope of the present manuscript we comment on the implications on the parameter space. The variations of $f_{\phi}$ and $\Lambda$ will shift the Planck normalization $a_i$ values in Fig. \ref{fig:numerical1}.
For certain choices of $f_{\phi}$ and $\Lambda$ we will not be able to satisfy the normalization for any $a_i$ values. However, the important point to note is that in the case where our normalization is satisfied at small $a_i$ values, the results will be similar to the vanilla curvaton which means negative $f_{NL}$. On the other hand for larger $a_i$ normalization, we should expect positive $f_{NL}$, $g_{NL}$ is however always positive irrespective of these choices.

\section{Conclusions and Discussion}

We investigated and showed that an axion like particle with the precise axion potential, coupled to a new confining dark sector (with matter content involving dark fermions) can generate the primordial density perturbation of our universe even if the same from the inflaton is not sufficient. The axion decay constant $f_{\phi}$, the scale of confinement $\Lambda$ of the dark sector and the masses $m_u$ and $m_d$ of the dark fermions that couple to the ALP are determined by observations in terms of the inflation scale $H_{\rm inf}$, which, in the minimal model, manifests a temporal de-confinement of the gauge group after inflation. Our investigations lead to the following results:
\begin{itemize}
\iffalse\item The axion-like scenario predicts a {\it positive}
local-type non-Gaussianity of order unity, $f_{\mathrm{NL}} \sim 1$,
on large scales.
This is in contrast to single-field inflation which yields a much
smaller local $f_{\mathrm{NL}}$, 
and also to a vanilla curvaton which produces
$f_{\mathrm{NL}} = -5/4 $ in the dominating limit ($R \to \infty$).
These values of non-Gaussianity are within reach of upcoming large-scale
structure surveys~\cite{Alvarez:2014vva}.\fi
\item For the pure cosine ALP potential we provide analytical estimates for the power spectrum, bi-spectrum and tri-spectrum as given in Eqns. \eqref{eq:Pzeta-fNL} \& \eqref{eq:gNL} and shown in Fig. \ref{fig:analytical} for the the parameter values $f_{\phi} = 2500 H_{\rm inf}$ and $\Lambda = 30 H_{\rm inf}$ when $m_u / m_d = 1$. Particularly the tri-spectrum for ALP is derived for the first-time in the literature according to our knowledge.
\item We studied the scenario for various combinations of $m_u$ and $m_d$ values which generates a non-perturbative potential which is not of purely cosine form (see Eqn. \eqref{main-pot}). We have derived the analytical expressions for the power spectrum and higher order correlation functions (see Eqns. \eqref{analytical-Ps}, \eqref{analytical-fNL} and \eqref{analytical-gNL})  for
$m_u/m_d = 1$ limit.  
 In this scenario we have found that the power spectrum behaves very similarly to the vanilla curvaton (with quadratic potential) scenario. But when we compute the bispectrum and trispectrum we can clearly see that these higher order correlation functions actually have an $a_i$ dependence when
for vanilla curvaton case these correlation functions are constants. For other limits of $m_u$ and $m_d$ we have done numerical estimations and we have observed that their power spectrum gets normalized (due to Planck observation) at different values of $a_i$ and as a result the amplitude of bispectrum and trispectrum are different for different cases (see Fig. \ref{fig:numerical1}).

\item We observed that the value of $f_{NL}$ and $g_{NL}$ are dependent on the ratio of $m_u$ and 
$m_d$. Apart from the $m_u = m_d$ case we have seen that the value of $f_{NL}$ is positive for all the  other cases and $g_{NL}$ is always positive irrespective of the ratio between $m_u$ and 
$m_d$. From Fig. \ref{fig:numerical1} it is evident that the value of $f_{NL}$ increases as the ratio between $m_u$ and $m_d$ becomes smaller i,e as the behavior of the potential deviates from pure cosine one towards the vanilla one the $f_{NL}$ increases. The same conclusion can be drawn for the nature of $g_{NL}$ also.
\item The results of our curvaton analysis in the limit $m_u = m_d$ resembles that of the vanilla curvaton (quadratic potential) scenario while in the limit $m_u \gg m_d$ resembles axion (pure cosine) potential, that is without having any corrections from new dark sector fermions.
\end{itemize}

These interesting features along with non-gaussianities $f_{NL}$ and $g_{NL}$ can be further
verified observationally by using, for instance, ultracompact minihalos and generation of Primordial Blackholes
~\cite{Ando:2018nge} and secondary Gravitational Waves, including the recently detected signals in NanoGrav, as shown in
Refs.~\cite{Kawasaki:2021ycf}; such a study is beyond the current draft and will taken up in later publication. Thus we
dare to imagine particle physics models such as ALP which is otherwise motivated from the Strong CP problem may also be
responsible for large scale structure as we see today, thereby connecting the small scale nature of quantum particle 
theory of fundamental interactions to the observation of very large scale structures that we see in the universe.

\section*{Acknowledgement}
\medskip

Authors thank Alessio Notari for collaborating on early stages of the project and Dario Bettoni, Maximilian Berbig and Keisuke Inomata for careful reading of the manuscript and suggestions.

%\section*{Extra}
%\textcolor{red}{We may vary the following parameters: $m_0/H$, $H/f$, $\alpha$, $\theta_i$, $\epsilon=\frac{\dot{H}}{H^2}$ and $m_u/m_d$. We have a constraint $P_\zeta$ and $n_S$, and upper bounds on $P_r$, $f_{NL}$ and $g_{NL}$. }

\end{document}